\documentclass[12pt]{article}
\usepackage[dvips]{color}
\usepackage{epsfig}
\usepackage{amsmath}
\usepackage{graphicx}

\begin{document}
\title{\bf Screening Length of Rotating Heavy Meson from AdS/CFT}
\author{J. Sadeghi $^{a,}$\thanks{Email:
pouriya@ipm.ir}\hspace{1mm} and
S. Heshmatian $^{a}$\thanks{Email: s.heshmatian@umz.ac.ir}\\
$^a$ {\small {\em  Sciences Faculty, Department of Physics, Mazandaran University,}}\\
{\small {\em P .O .Box 47415-416, Babolsar, Iran}}} \maketitle
\begin{abstract}
In this paper we consider a quark-antiquark ($q\bar{q}$) pair which
can be interpreted as a meson in ${\mathcal{N}}$=4 SYM thermal
plasma. We assume that the string moves at speed $v$ and rotates
around its center of mass simultaneously. By using the AdS/CFT
correspondence, we obtain the momentum densities of the rotating string
and determine its motion for small angular velocities. Then in general case,
we calculate the screening length of $q\bar{q}$ pair
numerically and show that its velocity dependance is in consistent
with the well known formula $L_s T\sim (1-v^{2})^{1/4}$ in the
literature.
\end{abstract}
\section{Introduction}
The AdS/CFT correspondence [1-3] plays an important role in many
complicated problems in QCD at strong coupling. For example, one of
these problems is the motion of charged particles through a thermal
medium. Already the subject of a quark in thermal plasma at weak
coupling has been  studied well [4-10]. But QCD at the strong
coupling will be a hard problem, however the AdS/CFT correspondence
solves the most of these complicated problems. On the other hand in
AdS/CFT correspondence there is the relation between type IIB string
theory in $AdS_{5}\times S^{5}$ space and $\mathcal{N}$=4 super
Yang-Mills (SYM) gauge theory on the 4-dimensional boundary of
$AdS_{5}$ space, this correspondence will be a candidate for solving
these problem. In that case, instead of a quark in gauge theory, we
consider dual picture of quark which is an open string in AdS space.
By adding temperature to the medium in gauge theory we have a black
hole (black brane) in $AdS_{5}$ space. There are many interesting
works in this field, for example, energy loss of quark and drag
force on moving quark through ${\mathcal{N}}$=4 Super Yang-Mills
thermal plasma [11-18]. Recently, the same calculations are done for
${\mathcal{N}}$=2 supergravity thermal plasma [19, 20]. This subject
is important because the solutions of supergravity theory with
$\mathcal{N}$=4 and $\mathcal{N}$=8 supersymmetry may be reduced to
the solutions of $\mathcal{N}$=2 supergravity. In Ref.s [19, 20] we
found that the problem of drag force in $\mathcal{N}$=2 supergravity
thermal plasma at zero non - extremality parameter is corresponding
to $\mathcal{N}$=4 SYM plasma for heavy quark.\\
The most fascinating problem is to consider a $q\bar{q}$ pair which
may be interpreted as a meson. As we know the problem of celebrated
Regge behavior of the hadron spectra has been discussed in
literature [21]. Also the meson spectrum obtained so far and
reasonably describing experiment can be seen in the Ref. [22]. In
this Ref. we see that the angular momentum plays an important role
to obtain the meson spectrum (in case of correction). So, this give
us motivation to consider the rotation of meson.\\
Already, energy of the moving $q\bar{q}$ pair through
${\mathcal{N}}$=4 SYM plasma is studied in both rest frames of
plasma and $q\bar{q}$ pair, which relate to each other by a Lorentz
transformation [23, 24, 25]. Authors in [25] found that the
$q\bar{q}$ pair feels no drag force, so such a system is an ideal
system. Actually, the $q\bar{q}$ pair may have more degrees of
freedom such as the rotational motion around the center of mass and
oscillation along the connection axis. In the Ref.s [26, 27] the
description of quark-antiquark system instead single quark well
explained. Also the problem of spinning open string (meson) in
description of non-critical string/gauge duality [28] considered. In
that paper the relationship between the energy and angular momentum
of spinning open string for the Regge trajectory of mesons
in a QCD-like theory is studied [29, 30].\\
In this paper, we add a rotational motion to the moving $q\bar{q}$
pair and obtain the momentum densities  $\Pi_{X}^{1}$ and
$\Pi_{Y}^{1}$ for such system. We assume that $q\bar{q}$ pair moves
at a constant velocity $v$ along $X$ direction and rotates
simultaneously around its center of mass. In this paper we will
obtain effect of rotational motion to the drag force on $q\bar{q}$
pair. First we find most general solutions for momentum densities
and then consider infinitesimal angular velocity to obtain drag
force on a heavy $q\bar{q}$ pair with non-relativistic velocity. We
obtain analytical solutions for $x$ and $y$ components of string as
functions of $r$ and momentum densities. Our method in this paper
differs from Ref. [28,]. In Ref. [31] the authors considered a
rotating quark and calculated the drag force on it through the plasma.\\
The calculation of screening length is one of the complicated
problems in QCD, but it can be studied in a strongly coupled
${\mathcal{N}}$=4 YM plasma [47 - 50]. As we know, the screening
length is direction dependent so it is interesting to calculate the
screening length in our rotating configuration. In general case
(relativistic) the integrals can not be solved analytically, so we
use the numerical method suggested in Ref. [47] to determine the
screening length at certain velocities. Then we show that the well
known formula $L_s T\sim (1-v^{2})^{1/4}$ is valid in our
configuration.

\section{String Equation of Motion}
An open string with two masses at its endpoints can be considered as
a model to describe a $q\bar{q}$ meson. In this picture two
point-like masses are connected with a string. This configuration
illustrates the strong interaction between two quarks due to the
gluon field flux tube and describes the confinement mechanism in
QCD. This model can also be used to investigate the orbitally
excited states of mesons and baryons. In the classical level, these
states can be regarded as the rotation of $q\bar{q}$ system. It is
necessary to consider more complicated motions of string to describe
the radial excitations and even other hadron excited states in
addition to the rotation. Indeed we have a spinning open string and
this picture is interesting because it is to the rotating meson.\\
\begin{tabular*}{2cm}{cc}
\hspace{0.50cm}\includegraphics[scale=0.5]{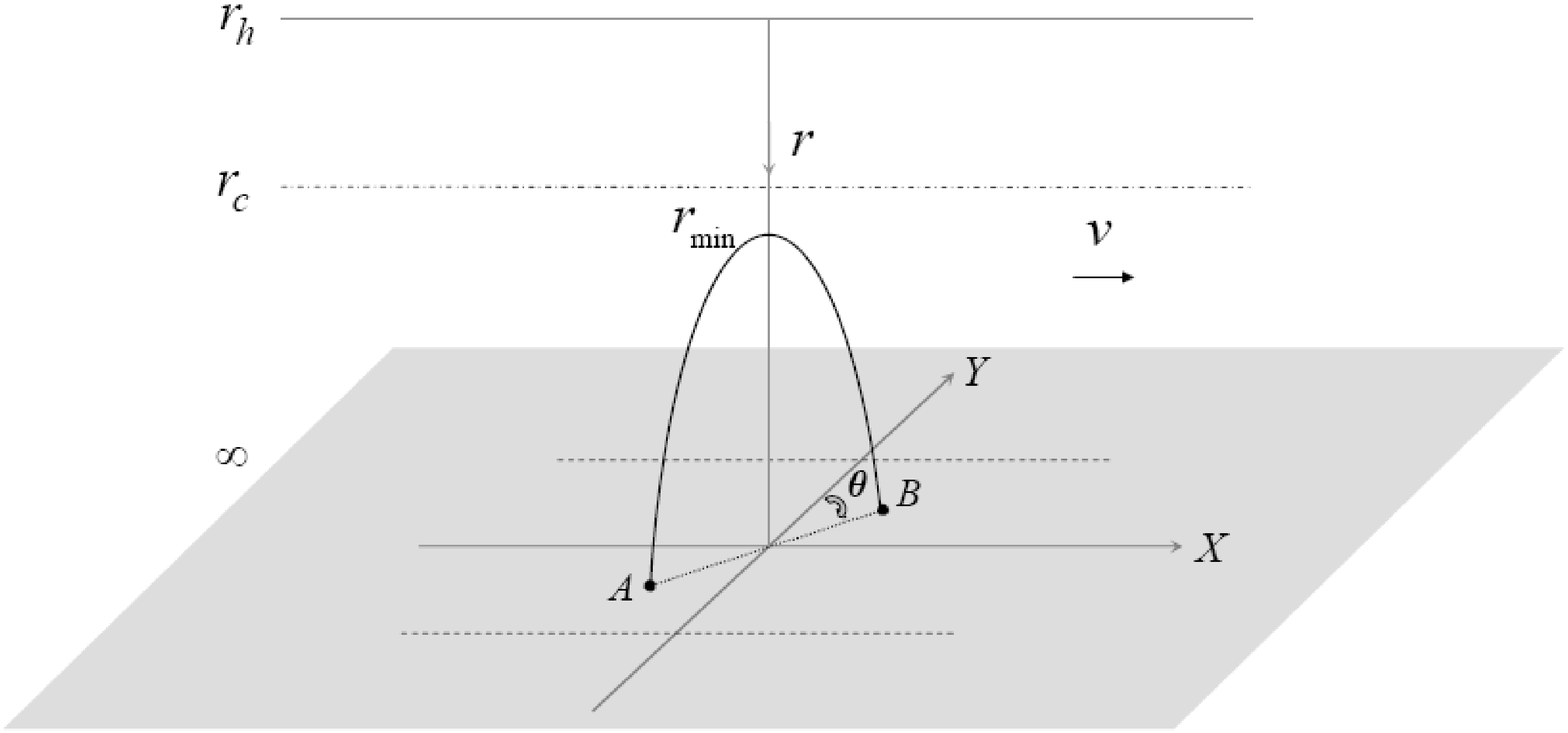}\\
\end{tabular*}\\
Figure 1: A rotating $\cap$ - shape string dual to a $q\bar{q}$ pair
which can be interpreted as a meson. $A$ and $B$ represent quark and
antiquark with separating length $L$. The radial coordinate $r$
varies from $r_h$ ( black hole horizon radius) to $r=r_{0}$ on
$D$-brane. $r_c$ is a critical radius, obtained for single quark
solution, which the string can not penetrate  and $r_{min}\geq r_c$.
$r_{min}=r_c$ is satisfied if $A$ and $B$ are located at origin
($L=0$), in that case there is straight string which is the dual
picture of a single static quark. $\theta$ is assumed to be the
angle with $Y$ axis and the string center of mass moves along $X$
axis at constant velocity $v$.\\\\\\\\\\\\\\
In order to represent the quark-antiquark pair, we consider an open
string in $AdS$ space and is stretched from $D$-brane to the black
hole horizon. The endpoints of string lie on $D$-brane and represent
quark and antiquark. The string starts from any point on $D$-brane
in $X$-$Y$ plane with radius $r=r_{0}$ to $r=r_{min}$ and then
returns to the other point on $D$-brane. We suppose the initial
condition that at $t=0$ the string is upright and its two endpoints
move with speed $v$ along $X$ axis and rotate around their center of
mass in $X$-$Y$ plane. The sketch of this string has been shown in Fig. 1.\\
According to the symmetry, we have $Y(r_{min})=0$ at $t=0$ without
the rotational motion and two halves of the string are attached
together smoothly. The string doesn't lean backward because the drag
force along the motion will be zero at $r=r_{min}$ due to the
symmetry. On the other hand, the force along the string should be
constant because each segment of string moves at constant velocity,
so we conclude that string remains upright [25]. Also there is the
Neumann boundary condition $X^{\prime}(r_{o})=0$ on D-brane. At the
presence of rotational motion, $F_{X}\propto \Pi_{X}^{1}$ and
$F_{Y}\propto \Pi_{Y}^{1}$ are constant with respect to $r$. That is
a consequence of the equation of motion for $X$ and $Y$, therefore the string remains upright.\\
From Maldacena dictionary we know that adding temperature to the
system is equal to the existence of a black hole at the center of
$AdS$ space. For the dual picture of ${\mathcal{N}}=4$ SYM plasma
there is the $AdS_{5}$ black hole solution which is given by [25],
\begin{eqnarray}\label{s1}
ds^{2}&=&\frac{1}{\sqrt{H}}(-hdt^{2}+d\vec{x}^{2})+\frac{\sqrt{H}}{h}dr^{2},\nonumber\\
h&=&1-\frac{r_{h}^{4}}{r^{4}},\nonumber\\
H&=&\frac{R^{4}}{r^{4}},
\end{eqnarray}
where $R$ is the curvature radius of AdS space and $r_{h}$ is the
radius of black hole horizon and $\vec{x}: (X, Y, Z)$. The string motion in considered to be in X-Y plan $Z=0$).\\
We know that the dynamics of an open string is described by the
Nambu-Goto action,
\begin{equation}\label{s2}
S=T_{0}\int{dtdr\mathcal{L}},
\end{equation}
where we used static gauge ($\sigma=r$ and $\tau=t$).
Therefore the lagrangian density of system may be
found as,
\begin{eqnarray}\label{s3}
{\mathcal{L}}&=&-\sqrt{-g}\nonumber\\
&=&-\left[1+\frac{h}{H}({X^{\prime}}^{2}+{Y^{\prime}}^{2})-\frac{1}{h}(\dot{X}^{2}+\dot{Y}^{2})-\frac{1}{H}(\dot{X}^{2}{Y^{\prime}}^{2}+\dot{Y}^{2}{X^{\prime}}^{2}-2\dot{X}X^{\prime}\dot{Y}Y^{\prime})\right]^{\frac{1}{2}},
\end{eqnarray}
where prime and dot denote derivative with respect to $r$ and $t$
respectively. Also $g\equiv det g_{ab}$ where $g_{ab}$ is the metric
on the string world sheet. It is the most general lagrangian for the
string with the endpoints on $D$-brane. We see that the lagrangian
density (3) depends on the derivatives of $X$ and $Y$, therefore
$\frac{\partial {\mathcal{L}}}{\partial X}=\frac{\partial
{\mathcal{L}}}{\partial Y}=0$. Then by using the lagrangian density
(3) the string equations of motion are given by the following
relations,
\begin{equation}\label{s4}
\frac{\partial}{\partial r}\left[\frac{1}{H\sqrt{-g}}((\dot{Y}^{2}-h)X^{\prime}-\dot{X}\dot{Y}Y^{\prime})\right]+\frac{1}{H}\frac{\partial}{\partial t}\left[\frac{1}{\sqrt{-g}}((\frac{H}{h}+{Y^{\prime}}^{2})\dot{X}-\dot{Y}Y^{\prime}X^{\prime})\right]=0,
\end{equation}
and
\begin{equation}\label{s5}
\frac{\partial}{\partial r}\left[\frac{1}{H\sqrt{-g}}((\dot{X}^{2}-h)Y^{\prime}-\dot{X}\dot{Y}X^{\prime})\right]+\frac{1}{H}\frac{\partial}{\partial t}\left[\frac{1}{\sqrt{-g}}((\frac{H}{h}+{X^{\prime}}^{2})\dot{Y}-\dot{X}X^{\prime}Y^{\prime})\right]=0,
\end{equation}
corresponding to $X$ and $Y$ respectively. In order to obtain the
total energy and momentum, drag force or energy loss of meson in the
thermal plasma, one should first calculate the canonical momentum
densities using the following expressions,
\begin{eqnarray}\label{s6}
\left(\begin{array}{ccc}
\Pi_{X}^{0} & \Pi_{X}^{1}\\
\Pi_{Y}^{0} & \Pi_{Y}^{1}\\
\Pi_{r}^{0} & \Pi_{r}^{1}\\
\Pi_{t}^{0} & \Pi_{t}^{1}\\
\end{array}\right)=-\frac{T_{0}}{H\sqrt{-g}} \left(\begin{array}{ccc}
\dot{Y}Y^{\prime}X^{\prime}-\dot{X}{Y^{\prime}}^{2}-\frac{H}{h}\dot{X} & \dot{X}\dot{Y}Y^{\prime}-X^{\prime}{\dot{Y}}^{2}+hX^{\prime}\\
\dot{X}X^{\prime}Y^{\prime}-\dot{Y}{X^{\prime}}^{2}-\frac{H}{h}\dot{Y} & \dot{X}\dot{Y}X^{\prime}-Y^{\prime}{\dot{X}}^{2}+hY^{\prime}\\
\frac{H}{h}(\dot{X}X^{\prime}+\dot{Y}Y^{\prime}) & -\frac{H}{h}({\dot{X}}^{2}+{\dot{Y}}^{2}-h)\\
h({X^{\prime}}^{2}+{Y^{\prime}}^{2}+\frac{H}{h}) & -h(\dot{X}X^{\prime}+\dot{Y}Y^{\prime})\\
\end{array}\right).
\end{eqnarray}
If we consider a  time dependent rotational motion, in
contradistinction to the previous works (without rotation) [12, 13,
25], the time derivative in the second term of equations (4) and (5)
doesn't vanish, so we have a complicated differential equation. But
we don't like to solve these equations and it is not the purpose of
this paper. Our aim is to determine the momentum densities of the
string and obtain the effect of rotational motion on the energy loss
and drag force. In the next section we consider a different method
with Ref.s [28, 31] to study the rotational motion.
\section{Rotating String}
We consider the following ansatz for the string,
\begin{eqnarray}\label{s7}
X(r, t)&=& vt+x\sin\omega t,\nonumber\\
Y(r, t)&=& y\cos\omega t,
\end{eqnarray}
where $v$ and $\omega$ are the linear and angular velocities
respectively. we choose $\theta(t)=\omega t$ as the angle with $Y$
axis as shown in Fig. 1. The functions $x$ and $y$ in the right hand
of equation (7) are only depend on $r$. Our goal in this paper is to
specify the motion of string and calculate the drag force on
$q\bar{q}$ pair. Then we calculate the screening length by using the
numerical method. We follow the methods of [12, 13, 19, 20, 25, 47]
and put $\dot{X}=v+\omega x \cos\omega t$,
$X^{\prime}=x^{\prime}\sin\omega t$, $\dot{Y}=-\omega y\sin\omega t$
and $Y^{\prime}=y^{\prime}\cos\omega t$ in the above equations. By
using equations (4), (5) and (6) one can obtain the following
equation,
\begin{equation}\label{s8}
\frac{\partial}{\partial t}(\Pi_{X}^{0}+\Pi_{Y}^{0})+
\frac{\partial}{\partial r}(\Pi_{X}^{1}+\Pi_{Y}^{1})=0,
\end{equation}
where,
\begin{eqnarray}\label{s9}
&&\left(\begin{array}{ccc}
\Pi_{X}^{0}\\
\Pi_{Y}^{0}\\
\Pi_{X}^{1}\\
\Pi_{Y}^{1}\\
\end{array}\right)=-\frac{T_{0}}{H\sqrt{-g}}\times\nonumber\\
&&\left(\begin{array}{ccc}
-x{y^{\prime}}^{2}\omega\cos^{3}\omega t-x^{\prime}y^{\prime}y\omega\cos\omega t\sin^{2}\omega t-{y^{\prime}}^{2}\cos^{2}\omega t-\frac{H}{h}(v+x\omega\cos\omega t)\\
y{x^{\prime}}^{2}\omega\sin^{3}\omega t+(v+x\omega\cos\omega t)x^{\prime}y^{\prime}\sin\omega t\cos\omega t+\frac{H}{h}y\omega\sin\omega t\\
x^{\prime}y^{2}\omega^{2}\sin^{3}\omega t-(v+x\omega\cos\omega t)yy^{\prime}\omega\sin\omega t\cos\omega t+hx^{\prime}\sin\omega t\\
hy^{\prime}\cos\omega t-y^{\prime}(v+x\omega\cos\omega t)^{2}\cos\omega t-(v+x\omega\cos\omega t)yx^{\prime}\omega \sin^{2}\omega t\\
\end{array}\right),
\end{eqnarray}
and,
\begin{eqnarray}\label{s10}
-g&=&1+\frac{h}{H}({x^{\prime}}^{2}\sin^{2}\omega t+{y^{\prime}}^{2}\cos^{2}\omega t)\nonumber\\
&-&\frac{1}{h}(v^{2}+x^{2}\omega^{2}\cos^{2}\omega t+2vx\omega \cos\omega t+y^{2}\omega^{2}\sin^{2}\omega t)\nonumber\\
&-&\frac{1}{H}({y^{\prime}}^{2}\cos^{2}\omega t(v+x\omega\cos\omega t)^{2}+{x^{\prime}}^{2}y^{2}\omega^{2}\sin^{4}\omega t)\nonumber\\
&-&\frac{2}{H}(v+x\omega\cos\omega t)x^{\prime}y^{\prime}y\omega\cos\omega t\sin^{2}\omega t.
\end{eqnarray}
In order to obtain the total energy and momentum of the string one
can use the following relations,
\begin{eqnarray}\label{s11}
E&=&-\int_{r_{min}}^{r_{0}} dr\Pi_{t}^{0},\nonumber \\
P_{X}&=&\int_{r_{min}}^{r_{0}} dr\Pi_{X}^{0},\nonumber\\
P_{Y}&=&\int_{r_{min}}^{r_{0}} dr\Pi_{Y}^{0},
\end{eqnarray}
where the energy density is given by,
\begin{equation}\label{s12}
\Pi_{t}^{0}=-\frac{hT_{0}}{H\sqrt{-g}}\left[{x^{\prime}}^{2}\sin^{2}\omega t+{y^{\prime}}^{2}\cos^{2}\omega t+\frac{H}{h}\right],
\end{equation}
and the momentum densities $\Pi_{X}^{0}$ and $\Pi_{Y}^{0}$ are given
by equation (9). After determining $x$ and $y$ one can obtain the
angular momentum of string using the following relation,
\begin{equation}\label{s13}
J=\int_{r_{min}}^{r_{0}} dr(X\Pi_{Y}^{0}-Y\Pi_{X}^{0}).
\end{equation}
In addition,  it is possible to study Regge trajectory by
calculating $\frac{E^{2}}{J}$ to specify the drag force. In that
case we should determine the string motion and it means that we
should find explicit expressions for $x(r)$ and $y(r)$. The
corresponding equations are complicated, so we consider a heavy
meson with small angular velocity to simplify them.
\section{Momentum Densities}
In this section we consider the special case of small angular
velocities to find explicit expression for $x$, $y$ and momentum
densities, then we can find drag force on $q\bar{q}$ pair. Also we
calculate the screening length of the rotating  $q\bar{q}$ pair
numerically. Here we consider a moving heavy quark-antiquark pair
which moves at constant velocity $v$ along $X$ axis and rotates
simultaneously around its center of mass. The angle with $Y$ axis
$\theta=\omega t$ and the angular velocity $\omega\ll 1$. This
assumption corresponds to the motion of a heavy meson with large
spin. Indeed in the very large angular momentum limit a
semiclassical approximation is valid. In this limit we can write,
\begin{equation}\label{s14}
\sqrt{-g}\approx\left[1-\frac{v^{2}}{h}+\frac{h}{H}
{x^{\prime}}^{2}\sin^{2}\omega
t+\frac{h-v^{2}}{H}{y^{\prime}}^{2}\cos^{2}\omega
t\right]^{\frac{1}{2}},
\end{equation}
and from equation (9) one find the following expression for the
momentum currents,
\begin{eqnarray}\label{s15}
\left(\begin{array}{ccc}
\Pi_{X}^{1}\\
\Pi_{Y}^{1}\\
\end{array}\right)=-\frac{T_{0}}{H\sqrt{-g}}
\left(\begin{array}{ccc}
hx^{\prime}\sin\omega t\\
(h-v^{2})y^{\prime}\cos\omega t\\
\end{array}\right),
\end{eqnarray}
where $\sqrt{-g}$ is given by (14). By using equations (14) and (15)
we can obtain the following equations,
\begin{eqnarray}\label{s16}
x^{\prime}&=&\frac{H(h-v^{2})}{h\sin\omega t}\Pi_{X}^{1}\sqrt{\frac{1}{(h-v^{2})(hT_{0}^{2}-H{\Pi_{X}^{1}}^{2})-hH{\Pi_{Y}^{1}}^{2}}},\nonumber\\
y^{\prime}&=&\frac{H\Pi_{Y}^{1}}{\cos\omega
t}\sqrt{\frac{1}{(h-v^{2})(hT_{0}^{2}-H{\Pi_{X}^{1}}^{2})-hH{\Pi_{Y}^{1}}^{2}}}.
\end{eqnarray}
It is clear that equations (16) are only depend on $r$ and $t$. If
we don't use the  small angular velocity assumption for heavy meson,
then the above solutions will depend on $r$, $t$, $x$ and $y$. From
[12, 13, 16] we learn that above expressions must be real, this
condition for single quark solution [12, 13, 16, 19, 20] yield to
the velocity-dependent critical radius,
\begin{equation}\label{s17}
r_{c}=\frac{r_{h}}{(1-v^{2})^{\frac{1}{4}}}.
\end{equation}
By using the reality condition in equation (16) for the
quark-antiquark system one can find special radius, $r_{min}$, where
functions $x$ and $y$ are not imaginary, thus string has real
energy. by using square root quantity in (16) one can obtain the
following relation for the turning point,
\begin{equation}\label{s18}
r_{min}=\left[r_{h}^{4}+\frac{1}{2T_{0}^{2}(1-v^{2})}\left(b+\sqrt{b^{2}-4T_{0}^{2}(1-v^{2})v^{2}r_{h}^{4}R^{4}{\Pi_{X}^{1}}^{2}}\right)\right]^{\frac{1}{4}},
\end{equation}
where we define,
\begin{equation}\label{s19}
b=R^{4}(1-v^{2}){\Pi_{X}^{1}}^{2}+T_{0}^{2}v^{2}r_{h}^{4}+R^{4}{\Pi_{Y}^{1}}^{2}.
\end{equation}
It is easy to check that $r_{min}\geq r_{c}$. If we consider
$\Pi_{X}^{1}=0$ ($L=0$), the special case of $r_{min}= r_{c}$ will
be satisfied. We note that $r_{min}= r_{c}$ corresponds to the
single quark solution [25].\\
Also we have the following useful conditions at $r=r_{min}$,
\begin{equation}\label{s20}
\frac{y^{\prime}}{x^{\prime}}=\cot\omega t.
\end{equation}
It is easily found that at $\omega\rightarrow0$ limit, the condition
(20) reduces to $\frac{y^{\prime}}{x^{\prime}}\rightarrow\infty$
[25]. Moreover from solution (7) we can see the boundary conditions
$X(r_{0}, t)=vt\pm{\frac{l}{2}}\sin\omega t$ and $Y(r_{0}, t)=\pm
\frac{l}{2}\cos\omega t$ which reduce to previous conditions without
rotational motion, namely $X(r_{0}, t)=vt$ and $Y(r_{0}, t)=\pm
\frac{l}{2}$for $\omega\rightarrow 0$. These boundary conditions can
also be satisfied with two separated string which move at velocity
$v$ along $X$ axis and simultaneously swing a circle of radius $\frac{l}{2}$.\\
Specifying these boundary conditions doesn't lead to a unique
solution for equation of motion, so we should specify additional
conditions for this motion. Here we assume that the string is
initially upright, move at velocity $v$ and rotates around its center of mass,
so it doesn't need any external agent to continue its motion.\\
Therefore by using relation (16), one can obtain the following
relation between momentum flows at $r=r_{min}$,
\begin{equation}\label{s21}
\Pi_{X}^{1}=\frac{h(r_{min})\Pi_{Y}^{1}}{h(r_{min})-v^{2}}\tan^{2}\omega
t.
\end{equation}
Therefore one can set momentum currents as following,
\begin{eqnarray}\label{s22}
\Pi_{X}^{1}&=&\frac{h(r_{min})\tan^{2}\omega t}{\sqrt{H(r_{min})(h(r_{min})-v^{2})+\frac{h(r_{min})\tan^{4}\omega t}{H(r_{min})}}},\nonumber\\
\Pi_{Y}^{1}&=&\sqrt{\frac{h(r_{min})-v^{2}}{H(r_{min})+\frac{h(r_{min})\tan^{4}\omega t}{H(r_{min})(h(r_{min})-v^{2})}}}.
\end{eqnarray}
As we see the $\omega=0$ limit leads us to have $\Pi_{X}^{1}=0$ and
$\Pi_{Y}^{1}=\sqrt{\frac{h(r_{min})-v^{2}}{H(r_{min})}}$ which is in
agreement with the results of Ref. [25]. Now we find that at the
presence of rotational motion, there is a non-zero $\Pi_{X}^{1}$ and
the value of $\Pi_{X}^{1}$ increases by $\omega$. On the other hand
by increasing $\omega$, the value of $\Pi_{Y}^{1}$ decreases. The
maximum value of $\Pi_{X}^{1}$ obtained at $\omega
t\rightarrow\frac{\pi}{2}$ limit, so one can write
$\Pi_{X}^{1}\approx\sqrt{\frac{r_{min}^{4}-r_{h}^{4}}{L^{4}}}$ and
$\Pi_{Y}^{1}=0$. Using equations (14) and (16) one can obtain
lagrangian density as following,
\begin{equation}\label{s23}
{\mathcal{L}}=-\sqrt{\frac{h-v^{2}}{h}}
\left[1+\frac{h}{H}\frac{H(h-v^{2}){\Pi_{X}^{1}}^{2}+{\Pi_{Y}^{1}}^{2}}{(h-v^{2})(hT_{0}^{2}
-H{\Pi_{X}^{1}}^{2})-hH{\Pi_{Y}^{1}}^{2}}\right]^{\frac{1}{2}},
\end{equation}
which reduces to the following expression at $v^{2}\rightarrow0$
limit (non-relativistic case),
\begin{equation}\label{s24}
{\mathcal{L}}=-
\left[1+\frac{1}{H}\frac{H(r^{4}-r_{h}^{4}){\Pi_{X}^{1}}^{2}+r^{4}{\Pi_{Y}^{1}}^{2}}{(r^{4}-r_{h}^{4})T_{0}^{2}
-R^{4}({\Pi_{X}^{1}}^{2}+{\Pi_{Y}^{1}}^{2})}\right]^{\frac{1}{2}}.
\end{equation}
In such limit, we can find analytical expressions for $x$ and $y$
from equation (16) as following,
\begin{eqnarray}\label{s25}
x&=&\frac{rR^{4}\,\mathcal{D}^{-\frac{1}{2}}}{r_{h}^{2}\,\sin\omega
t}\Pi_{X}^{1}\,\mathcal{A}\left[\frac{1}{4},\,\frac{1}{2},\,\frac{1}{2},\,\frac{5}{4},\,\frac{r^{4}\,T_{0}^{2}}{\mathcal{D}}\right],\nonumber\\
y&=&\frac{rR^{4}\,\mathcal{D}^{-\frac{1}{2}}}{r_{h}^{2}\,\cos\omega
t}\Pi_{Y}^{1}\,\mathcal{A}\left[\frac{1}{4},\,\frac{1}{2},\,\frac{1}{2},\,\frac{5}{4},\,\frac{r^{4}\,T_{0}^{2}}{\mathcal{D}}\right],
\end{eqnarray}
where
\begin{eqnarray}\label{s26}
\mathcal{D}&=&r_{h}^{4}\,T_{0}^{2}
+R^{4}({\Pi_{X}^{1}}^{2}+{\Pi_{Y}^{1}}^{2}),
\end{eqnarray}
and $\mathcal{A}$ is Appell function. Thus we success to determine
the motion of a non-relativistic meson in the ${\mathcal{N}}$=4 SYM
thermal plasma completely. In this case one can use equations (11)
and (13) to show that the ratio of squared energy to angular
momentum is proportional to string tension.\\
\section{Screening Length}
Now we are going to calculate the screening length for a
quark-antiquark pair with both linear and rotational motion. For screening length along $X$ and $Y$ we use
the following relations,
\begin{eqnarray}\label{s27}
L_{X}&=&2\,\int_{r_{min}}^{\infty} dr  X',
\nonumber\\
L_{Y}&=&2\,\int_{r_{min}}^{\infty} dr Y',
\end{eqnarray} \\
where $X'$ and $Y'$ are those in equation (16). In general case
($v^2\neq 0$) the integrals can not be solved analytically and we
should use the numerical method to solve them.\\
Here we follow the notation of Ref. [47] and change the variables
as,
\begin{equation}\label{s28}
\alpha=\frac{\Pi_{X}^{1}\, L^{2}}{r_{min}^{2}} , \, \, \beta =
\frac{\Pi_{Y}^{1} \, L^{2}}{r_{min}^{2}}, \,\, z=\frac{r}{r_{min}},
\,\, \rho=\frac{r_{h}}{r_{min}}.
\end{equation} \\
Applying these variables to equation (16) and use the hawking
temperature relation, we can write equation (27) in terms of new
variables and as following,
\begin{eqnarray}\label{s29}
L_{X}&=&2\,\int_{1}^{\infty} dz \,\, \frac{\alpha \,\rho
((1-v^{2})z^{4}-\rho^{4})}{\pi\,T\,r_{min}
\sin(\theta)\,z^{4}\,(z^{4}-\rho^{4})}\,\,\mathcal{F}(z) ,
\nonumber\\
L_{Y}&=&2\,\int_{1}^{\infty} dz \,\, \frac{\beta \,\rho
((1-v^{2})z^{4}-\rho^{4})}{\pi\,T\,r_{min}
\cos(\theta)\,z^{4}}\,\,\mathcal{F}(z) ,
\end{eqnarray} \\
where $\mathcal{F}(z)$ has the following form,
\begin{equation}\label{s30}
\mathcal{F}(z)=z^{4}\,
\left[T_{0}^{2}(z^{4}-\rho^{4})((1-v^{2})z^{4}-\rho^{4})+\rho^{4}(\alpha^{2}+\beta^{2})-z^{4}(\alpha^{2}+\beta^{2}-\alpha^{2}v^{2})\right]^{-\frac{1}{2}}\,.
\end{equation} \\
Now we are able to solve the two equations obtained in previous section simultaneously by the numerical
method suggested in [47]. First we choose $\theta=\pi/6$, then for
each value of $v$ in the range (0,1) we vary $\rho$ between 0 and 1.
We set $\alpha$ and $\beta$ in a way that the two integrals in
equation (28) have the same values. Finally, by inserting these
values of $\rho$, $\alpha$ and $\beta$ in equation (28) we obtain
the value of $LT$ and then plot $LT$ as a function of $\rho$. As we
expect, for a few values of $v$ this diagram has a peak which is
interpreted as the screening length $L_s$ of the rotating
$q\bar{q}$ pair, Figure (1).\\\\
\begin{tabular*}{2cm}{cc}
\hspace{1.3cm}\includegraphics[scale=0.60]{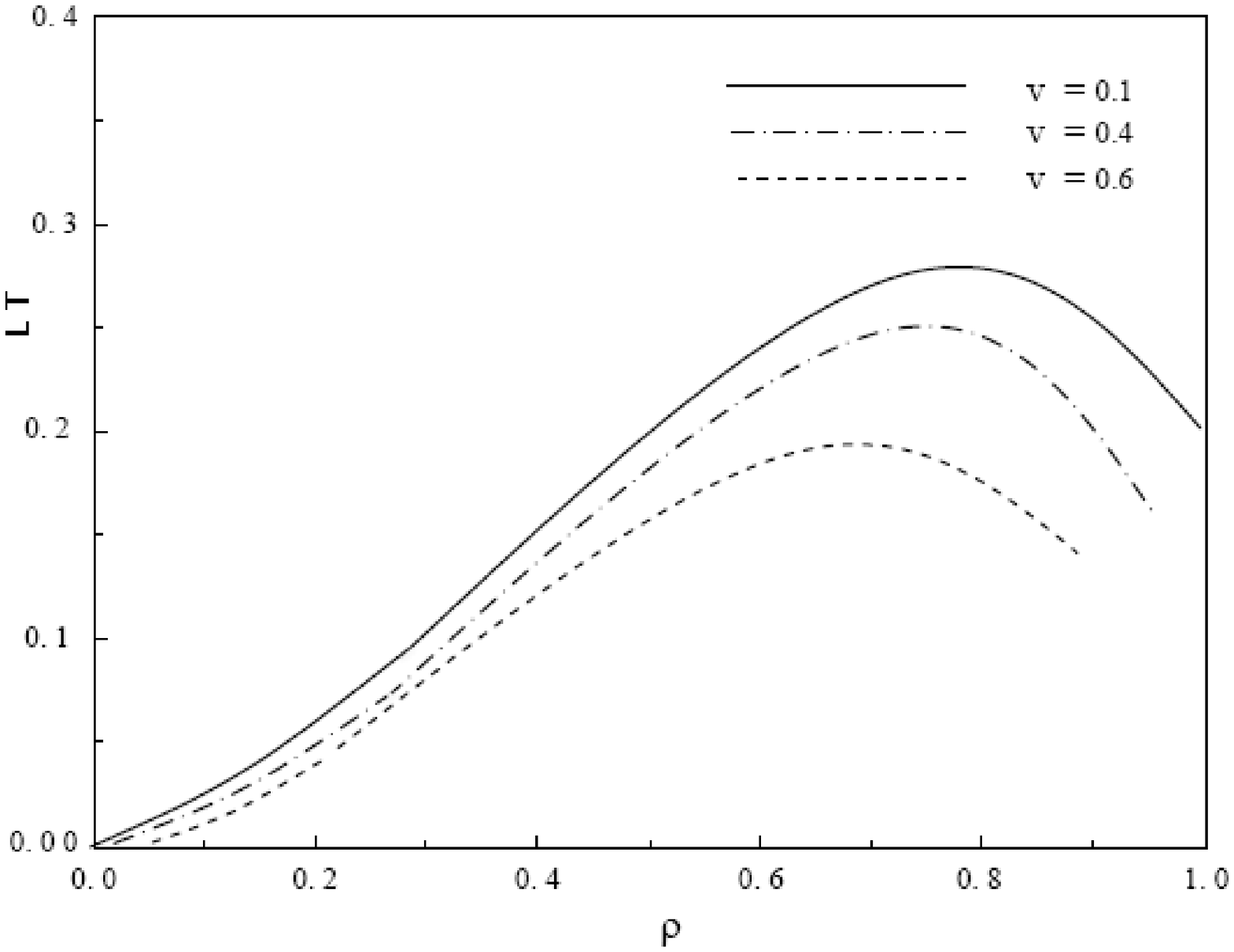}\\
\end{tabular*}\\
Figure 2: $LT$ vs. $\rho$ for some values of $v$. The maximum is
considered as screening length $L_s$. According to this diagram one
can find that as $v$ increases, the value of
$L_s$ gets smaller. \\\\\\\\\\
Here we demonstrate the velocity dependence of screening length .
From Fig. 2 it can be easily found that $L_s T\sim (1-v^2)^{1/4}$ is
valid. According to this graph as $v$ increases, $L_s$ decreases and
this result is in consistent with the corresponding formula. Here we
find a similar behavior for $LT$ vs. $\rho$ as in Ref. [47]. Our
result is also in a good agreement with the well known formula for
screening length and we find that $L_s T{(1-v^2)^{-\frac{1}{4}}}\simeq 1.1$ for the three various values of $v$ in Fig. 2.
\section{Conclusion}
This article is an extension of moving $q\bar{q}$ pair through
${\mathcal{N}}$=4 SYM thermal plasma which has done for the first
time (we follow different way with Ref.s [28, 31] and also the
calculation of the screening length for a rotating $q \bar{q}$ pair.
Already, the same problem for $v=0$ and $v\neq0$ is studied without
the rotation by using AdS/CFT correspondence[25]. Now we consider
the rotation of $q\bar{q}$ pair around their center of mass and
obtain the momentum flow along the string which are proportional to
the drag force. In the case of $\theta=\omega t$ at $\omega\ll 1$
limit for non-relativistic velocities we could specify the total
motion of the system. Without the rotation one can obtain the
momentum flows $\Pi$ proportional to the constant $C$ [12, 13, 19,
20, 25]. But in the case of rotational motion, we found the momentum
flows proportional to  $C\tan\omega t$ for small angular velocities.
We have shown that for the spinning string with very large angular
momentum, the value of momentum current along the string increases
in $X$ direction and decreases in $Y$ direction,
so the maximum value of $\Pi_{X}^{1}$ can be achieved for
$\omega t\rightarrow \frac{\pi}{2}$, where $\Pi_{Y}^{1}=0$.\\
Then we calculated the screening length for the considered
configuration numerically and checked that the $L_s T\sim
(1-v^{2})^{1/4}$ is valid for three various velocities. From Fig. 2
one can easily find that $L_s T{(1-v^{2})^{-\frac{1}{4}}}\simeq\,
1.1$ for the three various values of $v$. The value of $L_s T$ gets
smaller as $v$ increases and it is the result we
expected before.\\
Here, there are some interesting problem for future works. For
example one can obtain shear viscosity [32] or jet quenching
parameter [33-37] for rotating $q\bar{q}$ pair. Also it is
interesting to consider the effect of higher derivative terms as in
the previous cases [38-46]. As a recent work [47] one may consider
more quarks, such as four quarks in the baryon through
${\mathcal{N}}$=4 SYM thermal plasma. It may be interesting to
consider fluctuations of the quark-antiquark pair and obtain the
exact solution of such a system.

\end{document}